\documentclass[12pt]{article}

\def\gsim{\mathrel{\raise.4ex\hbox{$>$}\kern-0.8em\lower.7ex\hbox{$\sim$}}}
\def\lsim{\mathrel{\raise.4ex\hbox{$<$}\kern-0.8em\lower.7ex\hbox{$\sim$}}}

\usepackage{epsf}
\begin{document}

\begin{center}
{\Large  Stripe dynamics in presence of disorder and lattice potentials}

\smallskip

{C.\ Morais Smith$^{a,b}$, N.\ Hasselmann$^{a,c}$, Yu.\ A.\ Dimashko$^a$,
and A.\ H.\ Castro Neto$^c$ }

\smallskip

\noindent 
{\small
$^{a \,}$I Institut f{\"u}r Theoretische Physik, Universit{\"a}t 
Hamburg, D-20355 Hamburg, Germany \\
$^{b \,}$Institut de Physique Th{\'e}orique, Universit{\'e} de Fribourg, 
P{\'e}rolles, CH-1700 Fribourg, Switzerland \\
$^{c \,}$Dept. of Physics, University of California, Riverside, CA, 92521, 
USA \\
}
\end{center}
%\newpage

%\vspace*{2cm}

%\noindent C.\ Morais Smith, N.\ Hasselmann, Yu.\ A.\ Dimashko, and A.\ H.\ Castro Neto \\

%\vspace*{1.5cm}

\begin{center}
%\hspace*{0.8cm} 
\parbox{16.5cm}
{\footnotesize {\bf Abstract.}
We study the influence of disorder and lattice pinning on the dynamics of 
a charged stripe. Starting from a phenomenological model of a discrete
quantum string, we determine the phase diagram for this system. Three regimes 
are identified, the free phase, the flat phase pinned by the lattice, and the
disorder pinned phase. In the absence of disorder, the system
can be mapped onto a 1D array of Josephson junctions (JJ). 
The results are compared with measurements on nickelates
and cuprates and a good qualitative agreement is found
between our results and the experimental data.}  
\vspace{1cm}
\end{center}
\noindent {\bf 1. INTRODUCTION}\\

\noindent During the last years, there has been a great deal of interest concerning
the existence and dynamics of the striped phase in nickelate and cuprate 
materials. Theoretical [1], as well as experimental [2,3] investigations
have revealed the presence of charge and spin order.

In the present paper, we study within a phenomenological model the transversal
dynamics of a stripe pinned by an underlying periodic lattice and by a
random pinning potential provided by impurities. A phase diagram is determined 
as a function of the hopping parameter $t$, the stripe stiffness $J$ and the 
doping concentration $\nu$. At large values of $t/J$ and $\nu$, the stripe is
free and the system can be described by a quantum membrane. At low doping $\nu$,
the role played by the impurities is always relevant and the stripe is in a 
disorder pinned phase. When the effect of disordered impurities is negligible
and $t/J \ll 1$, the stripe is pinned by the underlying lattice. In this case,
by performing a dual transformation in the initial phenomenological Hamiltonian,
we can map the problem onto a 1D array of JJ and show that the quantum depinning
occurs at $(t/J)_c = 2/\pi^2$. Finally, we discuss our results and compare them
to recent experimental measurements on nickelates and cuprates.
\\

\noindent {\bf 2. THE MODEL}\\

\noindent 
Let us consider a single vertical stripe with one hole/site on a 2D lattice
in a disorder potential. The holes are allowed to hop only in the
transversal direction. The phenomenological Hamiltonian describing this
system is
\begin{equation}
\hat{H}=\sum_n \left[ - 2 t \cos \left( \frac{\hat{p}_n a}{\hbar}\right) + 
\frac{J}{2 a^2} \left(\hat{u}_{n+1}-\hat{u}_n \right)^2 + 
V_n (\hat{u}_n)\right].
\label{Ham1}
\end{equation}
Here, $t$ is the hopping parameter, $a$ is the lattice constant, $\hat{u}_n$ 
is the displacement of
the n-th hole from the equilibrium (vertical) configuration, $\hat{p}_n$
is its conjugate transversal momentum, $J$ is the stripe stiffness, and
$ V_n (\hat{u}_n)$ is an uncorrelated disorder potential, 
$< V_n (u) V_{n'} (u')>_D = D \delta (u - u') \delta_{n,n'}$, where
$<...>_D$ denotes the Gaussian average over the disorder ensemble and
$D$ is the inverse of the impurity scattering time.  

A dimensional estimate can provide us with the main features of the phase
diagram for this system. Let us start with the case of no impurity
potential $ V_n (\hat{u}_n)= 0$. 
%In this case, the pinning is only provided by the underlying lattice and 
%a typical displacement $\Delta u \sim a$. 
At large values of the hopping constant $t \gg J$, we can expand the
cos-term, $-2t\cos(\hat{p}_n a/\hbar) \sim const. + t (\hat{p}_n a/\hbar)^2$. 
The dynamics is then governed by the competition between the kinetic term
$t (k_n a)^2$ that tries to free the holes, and the elastic one, $(J/2a^2)
(\hat{u}_{n+1}-\hat{u}_n)^2$, that tends to keep the holes together. 
When the confinement is given by the lattice pinning potential, a typical
displacement $\hat{u}_{n+1}-\hat{u}_n \sim a$ and the wave vector 
$k_n \sim 1/a$. Hence, a transition from the flat phase, with the stripe pinned 
by the underlying lattice, to a free phase should occur at $t/J \sim 1$. 

Indeed, by performing a dual transformation in the Hamiltonian (\ref{Ham1})
to new variables referring to segments of the string, i.e., to a pair of
neighbour holes 
\begin{equation}
\hat{u}_n - \hat{u}_{n-1} = \hat{\pi}_n, \qquad 
\hat{p}_n = \hat{\varphi}_{n+1} - \hat{\varphi }_n.
\label{dt}
\end{equation}
and taking the limit of a large system, we obtain [4]
\begin{equation}
\hat{H}=-2t\sum_n\cos \left[ \frac{(\varphi_{n+1}-\varphi_n) a}{\hbar}\right]
- \frac{J}{2 a^2} \sum_n(\partial /\partial \varphi_n)^2, \\
\end{equation}
which is the Hamiltonian describing a Josephson junction chain (JJC).
The solution of this problem at $T = 0$ was found by Bradley and Doniach 
[5]. Depending on the ratio $t/J$, the  chain is either insulating 
(small $t/J$) or superconducting (large $t/J$). At the critical value
$(t/J)_c = 2 / \pi^2$, the JJC undergoes a Kosterlitz-Thouless like
superconductor/insulator transition. This transition is known to represent
the unbinding of vortex/antivortex pairs in the equivalent XY model. 

For the case of the stripe, this corresponds to a depinning transition
from the flat to the free phase, as we discussed above. The excitations
of the stripe are kinks (K) and antikinks (AK) and the transition in this case
corresponds to the unbinding of K/AK pairs. By exploiting the
relation of the stripe Hamiltonian to the sine-Gordon theory, we
calculated the infrared excitation spectrum of the quantum string and
showed that the insulating gap $\Delta \sim J$ present at $t = 0$
vanishes at $(t/J)_c = 2/\pi^2$ [4]. Hence, at large values of $t/J$ there is
a proliferation of kinks and antikinks and the striped structure
disappears. 

Let us now consider the other limit of strong pinning by impurities.
In this case, the potential provided by the lattice is irrelevant and
the typical displacement is of the order of the separation between
stripes $1/k_n \sim \hat{u}_{n+1}-\hat{u}_n \sim L$, see Fig.\ 1. By 
comparing now the kinetic
$t (a/L)^2$ and the elastic $J (L/a)^2$ terms, we observe that a
transition should occur at $t/J \sim (L/a)^4$. 

Indeed, by deriving the renormalization group (RG) differential equations
to lowest nonvanishing order in the lattice and disorder parameters, one
finds a set of flow equations [6], indicating that the transition
from the disorder pinned to the free phase occurs at $(t/J)_c = (18/\pi^2)
(L/a)^4$. The phase diagram of the striped phase is shown in Fig.\ 2, with
$\nu = a/L$. 

\vspace{6cm}

\noindent {\footnotesize {\bf Figure 1}: Striped phase. \hspace{5cm} {\bf Figure 2}:
Phase diagram.} \\ 

\noindent {\bf 3. DISCUSSIONS}\\

\noindent Next, we discuss the different phases displayed in Fig.\ 2 in the light
of experimental results obtained for the cuprates and nickelates. 
If neither disorder nor lattice potential are relevant, the stripe is
in the freely fluctuating Gaussian phase and the dynamics can be 
described by a quantum membrane. This is the case for doped La$_2$CuO$_4$,
that seems to be insensitive to disorder. The incommensurate (IC) spin 
fluctuations in La$_{2-x}$Sr$_x$CuO$_4$ [7] near optimal doping are 
strinkingly similar to those found in La$_2$CuO$_{4+\delta}$ [8], although
the oxygen doping produces annealed and the Sr doping produces quenched 
disorder. However, although weak disorder is unimportant near optimal doping,
we expect from Fig.\ 2  a critical doping $x_c \propto \nu_c$ below which 
disorder becomes relevant. Hence, for $\nu < \nu_c$ the stripes will be pinned, 
implying a broadening
of the IC spin fluctuations. Eventually, the IC peaks will overlap to produce
a single broad peak centered at the commensurate antiferromagnetic position.
This effect is actually observed in the spin glass phase $(x < 0.05)$ in neutron
scattering experiments [9], indicating the pinning of the stripes by disorder. 
For disorder pinned stripes, a depinning transition under strong magnetic fields 
has been predicted [10]. For a doping $x \sim 10^{-2}$ we estimate $E \sim 10^4$ V/cm.

Concerning the pinning provided by the underlying lattice, the very weak Bragg
peaks that are observable in La$_{2-x}$Sr$_x$CuO$_4$ only at $x = 1/8$ [11] indicate
that the striped phase is nearly always in a floating phase IC with the lattice.
However, by doping this material with Nd, one produces a tilt in the oxygen octahedra
in the CuO planes, increasing hence the pinning by the lattice. Indeed, 
La$_{1.6-x}$Nd$_{0.4}$Sr$_x$CuO$_4$ shows strong lattice commensurability effects.
The static stripe order is most pronounced at the commensurate $x \approx 1/8$ [2],
but has also been observed at other compositions [3]. In sum, the Nd doping pushes the
roughening transition to higher values of $t/J$ and drives the cuprate materials into the
flat phase.

Let us now consider the nickelates. Weak disorder is clearly relevant in these
materials, since the width of the IC peaks is always much narrower in the oxygen
doped samples [12-14] than in the Sr doped ones [15-17]. In addition, both, Sr and oxygen
doped nickelates show strong commensuration effects [12,17], indicating that the
striped phase couples strongly to the lattice. Therefore, we conclude that for the
nickelates both, pinning by the lattice and by impurities are relevant.

In conclusion, we studied the competition between disorder and lattice effects on
the transverse stripe fluctuations for nickelates and cuprates and identified the
different phases in a $t/J$ versus $\nu$ phase diagram. 
\\

\noindent{\bf Acknowledgement}\\

\noindent 
The authors gratefully acknowledge financial support by DAAD-CAPES Project No.
415-probral/sch\"u.\\

\noindent{\bf References}\\

\noindent
\begin{tabular}{rl}
\noindent 1.& H.\ Eskes \textit{et al.}, Phys.\ Rev.\ B \textbf{58}, 13265 (1998) and
references therein. \\
2.& J.\ M.\ Tranquada \textit{et al.}, Nature \textbf{375}, 561
(1995); Phys.\ Rev.\ B \textbf{54}, 7489 (1996). \\
3.& J.\ M.\ Tranquada \textit{et al.}, Phys.\ Rev.\ Lett. \textbf{%
78}, 338 (1997).\\
4.& J.\ Dimashko {\it et al.}
to appear in  Phys.\ Rev.\ B, July (1999). \\
5.&  R.\ M.\ Bradley and S.\ Doniach, Phys.\ Rev.\ B {\bf 30}, 1138 
(1984). \\ 
6.& N.\ Hasselmann {\it et al.}
Phys.\ Rev.\ Lett. {\bf 82}, 2135 (1999). \\
7.& G.\ Aeppli {\it et al.}, Science {\bf 278}, 1432 (1997); T.\ E.\ Mason
{\it et al.}, Phys.\ Rev.\ Lett. {\bf 68}, 1414 (1992). \\
8.& B.\ O.\ Wells {\it et al.}, Science {\bf 277}, 1067 (1997).\\
9.& K.\ Yamada {\it et al.}, Phys.\ Rev.\ B {\bf 57}, 6165 (1998).\\
10.&  C.\ Morais Smith \textit{et al.}, Phys.\ Rev.\ B \textbf{58}, 453 (1998).\\
11.& T.\ Suzuki {\it et al.}, Phys.\ Rev.\ B {\bf 57}, R3229 (1998).\\
12.& P.\ Wochner {\it et al.}, Phys.\ Rev.\ B {\bf 57}, 1066 (1998). \\
13.& K.\ Nakajima {\it et al.}, J.\ Phys.\ Soc.\ Jpn.\ {\bf 66}, 809 (1997). \\
14.& J.\ M.\ Tranquada \textit{et al.}, Phys.\ Rev.\ Lett. \textbf{%
73}, 1003 (1994); {\it ibid.}, {\bf 79}, 2133 (1997). \\
%12.& N.\ L.\ Saini {\it et al.},  Phys.\ Rev.\ Lett. {\bf 79}, 3467 (1997). \\
15.& G.\ Blumberg {\it et al.},  Phys.\ Rev.\ Lett. {\bf 80}, 564 (1998). \\
16.& S.\ H.\ Lee and S.-W.\ Cheong,  Phys.\ Rev.\ Lett. {\bf 79}, 2514 (1997). \\
17.& J.\ M.\ Tranquada \textit{et al.}, Phys.\ Rev.\ B \textbf{54}, 12 318 (1996). \\
\end{tabular}

\newpage

\begin{figure}[t]
\unitlength1cm
\vspace{0.1cm}
\begin{picture}(14,10)
\epsfxsize=10cm
\put(1.5,0.5){\epsfbox{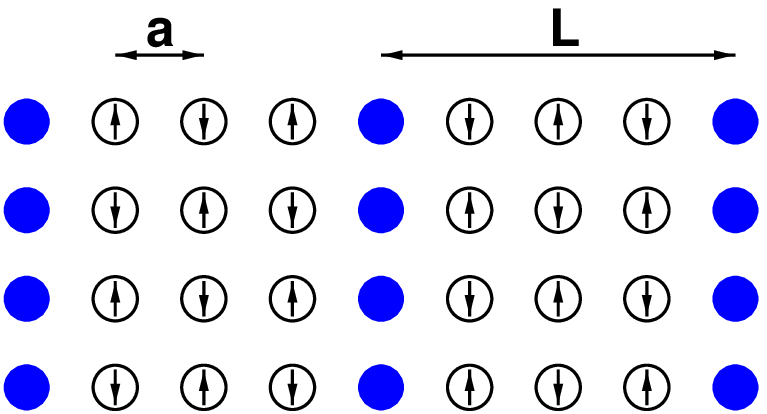}}
\end{picture}
\caption[]{}
%{\label{fig1}}
%
\end{figure}

\begin{figure}[t]
\unitlength1cm
\vspace{0.1cm}
\begin{picture}(14,10)
\epsfxsize=10cm
\put(1.5,0.5){\epsfbox{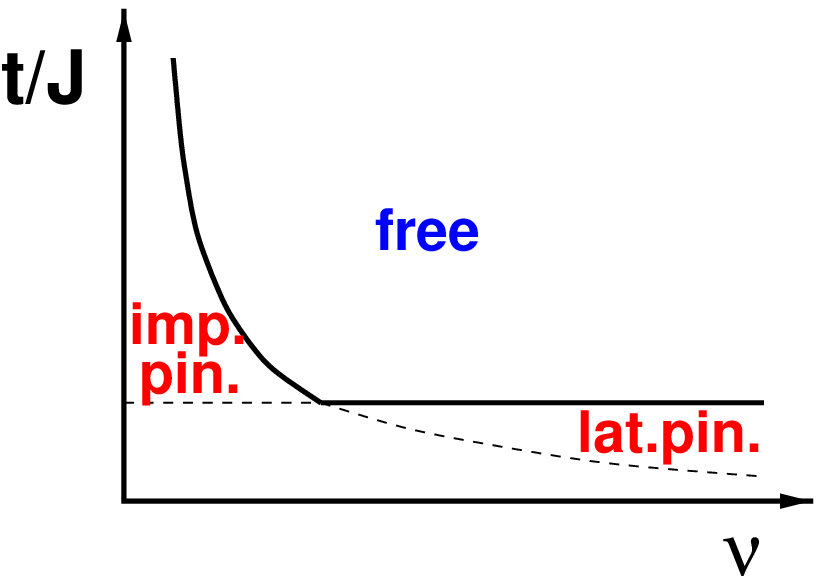}}
\end{picture}
\caption[]{}
%{\label{fig1}}
%
\end{figure}

\end{document}